\documentclass{aa}

\newcommand\Rafrac[4]{#1^{\rm h}#2^{\rm m}#3\fs #4}

\newcommand\teff{T_{\mathrm{eff}}}

\newcommand\loghe{\log \frac{n_{\mathrm{He}}}{n_{\mathrm{H}}}}

\newcommand{\Rsolar}{\mbox{\,$\rm R_{\odot}$}}        

\begin{document}
\title{Binaries discovered by the SPY project
I.\ HE\,1047--0436: a subdwarf\,B + white dwarf system%
\thanks{Based on data obtained at the Paranal Observatory of the European 
Southern Observatory for program No.\ 165.H-0588 and 266.D-5658}
\thanks{Based on observations collected at the German-Spanish Astronomical 
Center (DSAZ), Calar Alto, operated by the Max-Planck-Institut f\"ur 
Astronomie Heidelberg jointly with the Spanish National Commission for 
Astronomy} 
}
\author{R. Napiwotzki\inst{1}, H. Edelmann\inst{1}, 
U. Heber\inst{1}, C. Karl\inst{1}, H. Drechsel\inst{1}, 
E.-M. Pauli\inst{1}, N. Christlieb\inst{2} 
}
\offprints{R. Napiwotzki}
\institute {Dr. Remeis-Sternwarte, Astronomisches Institut der Universit\"at
Erlangen-N\"urnberg, Sternwartstr. 7, D-96049 Bamberg, Germany; 
e-mail: napiwotzki@sternwarte.uni-erlangen.de
\and
Hamburger Sternwarte, Universit\"at Hamburg, Gojenbergsweg 112, 
D-21029 Hamburg, Germany
}

\date{}

\abstract{In the course of our search for double degenerate binaries as 
potential progenitors of type Ia supernovae with the UVES spectrograph at 
the ESO VLT 
(ESO {\bf S}N\,Ia {\bf P}rogenitor surve{\bf Y} -- SPY)
we discovered that the sdB star HE\,1047$-$0436 is radial 
velocity variable.
The orbital period of 1.213253\,d, 
a semi-amplitude of 94\,km\,s$^{-1}$, and a minimum mass 
of the invisible companion of $0.44\,M_\odot$ are derived from the analysis
of the radial velocity curve. 
We use an upper limit on the projected rotational velocity of the sdB star
to constrain the system inclination and the companion mass to 
$M>0.71\,M_\odot$,
bringing the total mass of the system closer to the Chandrasekhar limit.
However, the system will merge due to loss of angular momentum via 
gravitational wave radiation only after several Hubble times. 
Atmospheric parameters and 
metal abundances are also derived.
The resulting values are typical for sdB stars.
\keywords{Stars: early-type -- binaries: spectroscopic -- Stars: fundamental
parameters -- white dwarfs}}

\authorrunning{Napiwotzki et al.}
\titlerunning{Binaries discovered by the SPY project. I.\
HE\,1047$-$0436}
\maketitle

\section{Introduction\label{intro}} 
 
Type Ia supernovae (SNe Ia) play a prominent role in the study of cosmic 
evolution.
In particular they are regarded as one of
the best standard candles
to determine 
the cosmological parameters $H_\circ$, $\Omega$ and $\Lambda$ 
(e.g.\ Riess et al.\ \cite{ries98}, Leibundgut \cite{L01}).  
What kind of stars produce SN~Ia events remains largely a mystery
(e.g. Livio \cite{liv00}). There is
general consensus that the event is due to the thermonuclear explosion of a 
white dwarf (WD) when a critical mass is reached, but the nature of the
progenitor system remains unclear.

One of the two viable scenarios 
is the so-called double degenerate (DD) scenario (Iben \& Tutukov 1984).
The DD model considers a binary with the sum of the masses of the WD
components larger than the Chandrasekhar mass, which merges in less than a 
Hubble time due the loss of angular momentum via gravitational wave 
radiation.
Previous radial velocity (RV) surveys have discovered about 
15 DDs with $P<6\fd 3$ (see Marsh \cite{mar00}). 
None of these systems seems massive enough to qualify as a SN\,Ia
precursor. 
In order to perform a definite test of the DD scenario we 
have embarked on a large 
spectroscopic survey of 1500 WDs using the UVES spectrograph at 
the ESO VLT UT2 (Kueyen) to search for RV variable WDs
and pre-WDs (ESO {\bf S}N\,Ia {\bf P}rogenitor 
surve{\bf Y} -- SPY).

Recently, subluminous B stars (sdB) with WD companions have been 
proposed as potential SNe\,Ia progenitors by Maxted et al.\ (\cite{mama00}) 
who discovered that KPD\,1930$+$2752 is a sdB+WD system.
Its total mass exceeds the critical
mass and the components will merge within a Hubble time 
which makes KPD\,1930$+$2752 the best candidate for a SN\,Ia progenitor 
known today (although this interpretation has been questioned recently, Ergma 
et al.\ \cite{erfe01}). 

SdB stars are pre-white dwarfs of low mass ($\approx$ $0.5\,M_\odot$) 
still burning helium in the core, which will evolve directly
to the WD stage omitting a second red giant phase 
(Heber \cite{heb86}).
Only seven sdB+WD systems with known periods are available in the 
literature. 
Except for KPD\,1930$+$2752 their total masses do not exceed the 
Chandrasekhar limit.

Here we report on follow-up spectroscopy of HE\,1047$-$0436 
(Section~\ref{obs}).
The RV curve is derived and the nature of the companion is
discussed in Section~\ref{rad}.
A spectroscopic analysis and further constraints\nopagebreak[3] on the
system inclination and companion mass are presented in Section~\ref{spec}.

\section{Observations and Data Analysis\label{obs}}

HE\,1047$-$0436 ($\alpha_{2000}= \Rafrac{10}{50}{26}{9}$, 
$\delta_{2000}=-4^{\circ} 52'36''$, 
$B_{\mathrm{pg}}= 14\fm 7$) was discovered by the 
Hamburg ESO survey (HES; Wisotzki et al.\ \cite{wis00}, 
Christlieb et al.\ \cite{chri01}) as a potential hot WD 
and, therefore, was included in our survey. 
The UVES spectra showed 
that it is in fact a sdB star (Christlieb et al.\ \cite{chri01}) and  
a rather large
RV shift of 160\,km/s was found 
which made the star a prime target for further study.

14 high resolution Echelle spectra of HE1047-0436 have 
been secured with VLT-UVES between March 7 and 
18, 2001. The nominal resolution with the $2.1''$ slit used by us
amounts to $\frac{\lambda}{\delta \lambda} = 19000$. Additional long slit 
spectra of somewhat lower resolution (1\,\AA) have been obtained at the Calar 
Alto observatory using the TWIN spectrograph on March 11 and
12, 2001. 
Details on the observational set up of the UVES instrument and the data 
reduction can be found in Koester et al.\ (\cite{koe01}). The TWIN 
instrument and the reduction strategy are described in Napiwotzki \&
Sch\"onberner (\cite{NS95}).

\section{Radial velocity curve and the nature of the invisible companion
\label{rad}}

Radial velocities from all UVES and TWIN spectra were derived by cross 
correlating parts of the blue spectra that display numerous sharp metal 
absorption lines besides helium and Balmer lines to a synthetic spectrum 
calculated from an LTE model atmosphere (see below). 
The measurements are
accurate to $\pm$3\,km~s$^{-1}$ for the UVES data and somewhat less precise 
($\pm$12\,km~s$^{-1}$) 
for the TWIN data because of the lower spectral resolution of the latter
instrument. 

\begin{figure}
\vspace*{6.3cm}
\includegraphics{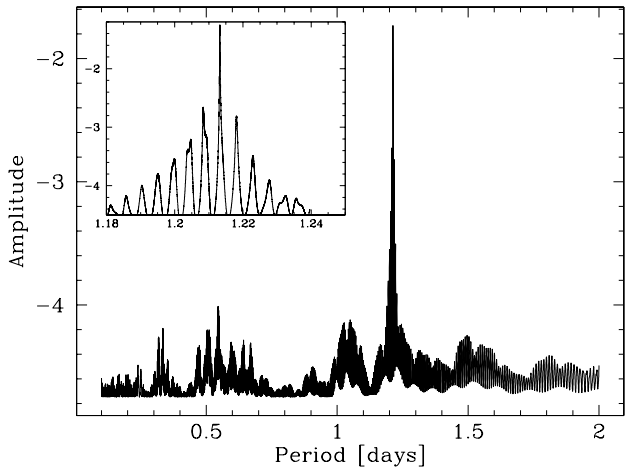}
\caption{Power spectrum of the HE\,1047$-$0436 measurements. The inset 
shows details of the region around the main peak
\label{power}}
\end{figure}

\begin{figure}
\vspace*{6.3cm}
\includegraphics{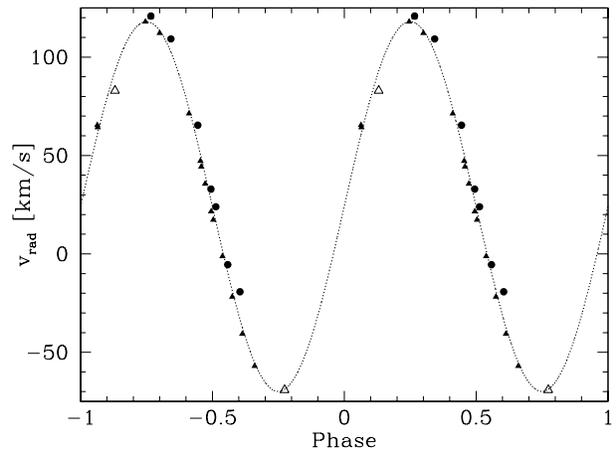}
\caption{Measured RVs as a function of orbital 
phase and fitted sine curve for HE\,1047$-$0436. Filled triangles indicate 
2001 UVES observations, open triangles both 2000 UVES discovery observations,
and filled circles 2001 TWIN measurements
\label{radvel}}
\end{figure}

Since this system is single-lined the analysis
is straightforward. 
A periodogram analysis was performed and the resulting 
power
spectrum is shown in Fig.~\ref{power} and the best fit 
RV curve in Fig.~\ref{radvel}. 
The discovery spectra taken at April 21 and May 17, 2000 are included in 
Fig.~\ref{power} 
and Fig.~\ref{radvel}. Since these spectra were taken about one year before 
the 2001 data a very accurate period could be determined:
$P = 29^{\mathrm{h}} 7^{\mathrm{m}} 5^{\mathrm{s}}$
and a semi-amplitude of 94\,km\,s$^{-1}$. The system velocity is
$\gamma_0 = 25\,\mathrm{km}\,\mathrm{s}^{-1}$ (this has to be
corrected by a gravitational redshift of 1.9$\,\mathrm{km}\,\mathrm{s}^{-1}$
to derive the real system velocity). 
Accordingly the ephemeris for the time $T_0$ defined as the conjunction time 
at which the star moves from the blue side to the red side of the RV curve 
(cf.\ Fig.~\ref{radvel}) is
\begin{displaymath}
\mathrm{Hel.JD} (T_0) = 244\,51975.03228 + 1.213253\times E
\end{displaymath}
Two aliases exist, which differ by 
$\pm 6^{\mathrm{m}}$, but they can be ruled out on a high confidence
level (cf.\ inset of Fig.~\ref{power}).

Since HE\,1047$-$0436 is a single-lined binary we can only
use the mass function to constrain the mass of the invisible companion.
From a comparison with evolutionary calculations (Dorman et al.\ \cite{DRC93})
we adopted
$M=0.474\,M_\odot$ for the sdB primary.
The mass function yields a lower limit of
$0.44\,M_\odot$ for the invisible companion (for an inclination
$i = 90^\circ$). 
Therefore we conclude that it is a WD with a C/O core.
A main sequence companion with such a high mass can be ruled out, because 
it would leave a detectable fingerprint in our high-resolution spectra.
In the next section we will show how the measurement of the projected
rotational velocity of the sdB constrains the inclination $i$ and the 
WD mass.

\section{Spectroscopic analysis\label{spec}}

\begin{figure}
\vspace*{10.3cm}
\includegraphics{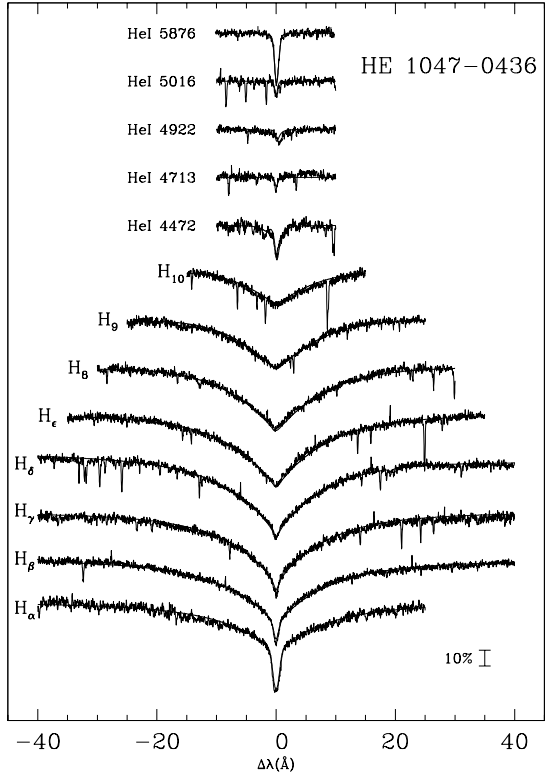}
\caption{Fit of the hydrogen and helium lines of the coadded spectrum of
HE\,1047$-$0436\label{fit}}
\end{figure}
\begin{figure}
\vspace*{5.0cm}
\includegraphics{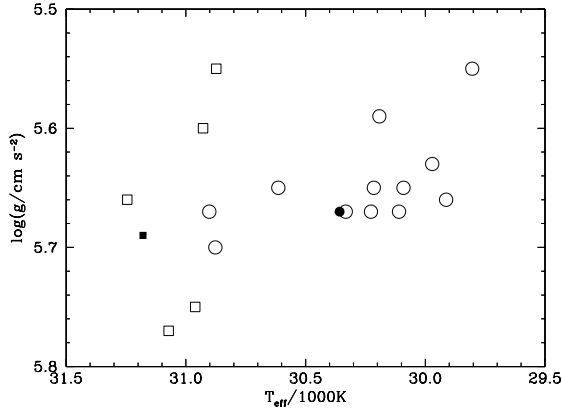}
\caption{Parameters derived from individual UVES and TWIN spectra (open 
circles and squares) and the coadded spectra (filled symbols)\label{teffg}}
\end{figure}

The stars in the HE\,1047$-$0436 binary are only
separated by $\approx 5 R_\odot$ (slightly depending on $i$). 
Thus the system must have gone through phases of 
heavy interaction, probably common envelope evolution, during the red
giant stages of the progenitors. Thus we checked if the sdB parameters and
chemical abundances deviate from normal sdBs.

In order to improve the S/N ratio the UVES spectra were RV 
corrected and then coadded.
The coadded as well as the individual UVES and spectra TWIN have been 
analyzed to derive atmospheric parameters using NLTE model atmospheres 
(Napiwotzki \cite{napi97}) 
and a $\chi^2$ procedure described by Napiwotzki et al.\ 
(\cite{napi99}). 
An LTE metal abundance analysis of the coadded UVES spectrum was performed 
(see Heber et al.\ \cite{hebe00} 
for details of the models and the fit procedure).

Matching the synthetic Balmer and He I line profiles to the observations 
resulted in effective temperature, gravity and He abundance with 
very small fitting errors, i.e. $T_{\mathrm{eff}} = 30242\pm39$\,K; 
$\log g = 5.66\pm0.008$ and $\loghe = -2.632\pm0.026$ 
(Fig.\ref{fit}). 
Adopting the sdB mass of $0.47\,M_\odot$ the radius can be calculated from 
the gravity: $R=0.17\pm 0.02\Rsolar$. 

Since more than a dozen individual UVES spectra are available systematic 
errors can be estimated. Individual results (with one exception) differ by no 
more than $\pm$500K and $\pm 0.05$\,dex in $\log g$ from that derived from 
the coadded 
spectrum (see Fig.\ref{teffg}). Another important accuracy check can be 
obtained from the long slit spectra. The fit of the coadded TWIN spectrum 
resulted in a moderately higher
effective temperature (about 1000\,K) 
than from the UVES spectrum, whereas the gravity is 
in very close agreement (cf.\ Fig.~\ref{teffg}).

Lines of several heavy elements 
can be identified
in the UVES spectra (cf.\ Table~\ref{abu}). 
The large number of \ion{N}{II} lines allows to determine the 
microturbulent velo\-city which turns out to be zero. 
The resulting abundances are summarized in 
Table~\ref{abu} and compared to the sdB star 
Feige\,36 (Edelmann et al.\ \cite{E99}) which has 
atmospheric parameters ($\teff=29700$\,K, $\log g=5.9$, $\loghe=-2.1$) very 
similar to HE\,1047$-$0436 and to solar 
composition.
In order to match the 
ionization equilibria of C, N, Si, and S it would be necessary to 
lower $\teff$ to 29\,000\,K. 
This is probably caused by the neglect of NLTE effects. However, the 
abundances are almost unaltered (by less than 0.1 dex) by such a change in 
$\teff$.  

The comparison of the metal abundances of HE\,1047$-$0436 and Feige\,36
in Table~\ref{abu} shows that the abundance patterns are quite similar.
Only iron is somewhat underabundant in HE\,1047$-$0436 (only an upper
limit could be derived). 
We conclude that the phases of close binary interaction didn't produce
any detectable peculiarities of the sdB star.

\begin{table}
\caption{Metal abundances for HE\,1047$-$0436 compared to 
Feige\,36 (Edelmann et al.\ \cite{E99}) and to solar composition
(Gre\-vesse \& Sauval \cite{grsa98}). $n$ is the number of spectral lines per 
ion}\label{abu}
\begin{tabular}{|l|rlll|}
\hline
ion            & $n$   & $\epsilon$(HE\,1047$-$0436)  
		& $\epsilon$(Feige\,36)   & $\epsilon$(sun)  \\
\hline
\ion{C}{II}    & 4   & 8.03$\pm$0.12   &                             & 8.52 \\
\ion{C}{III}   & 3   & 7.94$\pm$0.11   & 7.18$\pm$0.20               &  \\
\ion{N}{II}    & 49  & 8.01$\pm$0.19   & 7.67$\pm$0.12               & 7.92 \\
\ion{N}{III}   & 1   & 8.40            & 		             &  \\
\ion{O}{II}    & 23  & 7.80$\pm$0.11   & 7.94$\pm$0.17               & 8.83 \\
\ion{Mg}{II}   & 1   & 6.78            & 6.58                        & 7.58 \\
\ion{Al}{III}  & 3   & 5.60$\pm$0.03   & 5.75$\pm$0.06               & 6.47 \\
\ion{Si}{III}  & 7   & 6.82$\pm$0.23   & 6.93$\pm$0.18               & 7.55 \\
\ion{Si}{IV}   & 1   & 6.60            &                             &  \\
\ion{S}{II}    & 14  & 7.57$\pm$0.23   & 7.71$\pm$0.43               & 7.33 \\
\ion{S}{III}   & 11  & 7.48$\pm$0.33   & 7.34$\pm$0.47               &  \\
\ion{Ar}{II}   & 7   & 7.17$\pm$0.22   & 7.09$\pm$0.10               & 6.40 \\
\ion{Fe}{III}  &     & $<$6.8          & 7.20$\pm$0.26               & 7.50 \\
\hline
\end{tabular}
\end{table}

A closer inspection of the metal lines showed
that these are much narrower than expected from the nominal resolution
($R\approx 19000$). The obvious reason is that the seeing during all
exposures ($0.4''\ldots 1.2''$) was smaller than the slit width ($2.1''$).
This enables us to derive meaningful upper limits on the projected rotational
velocity $v \sin i$ of the sdB. Since we were interested in minimizing the
instrumental broadening we coadded only the seven spectra taken during 
periods of best seeing ($0.4''\ldots 0.8''$) for this task. We adopted
a conservative value of the spectral resolution corresponding to
the mean of the three best seeing values ($0.5''$). 

\begin{figure}
\vspace*{4.8cm}
\includegraphics{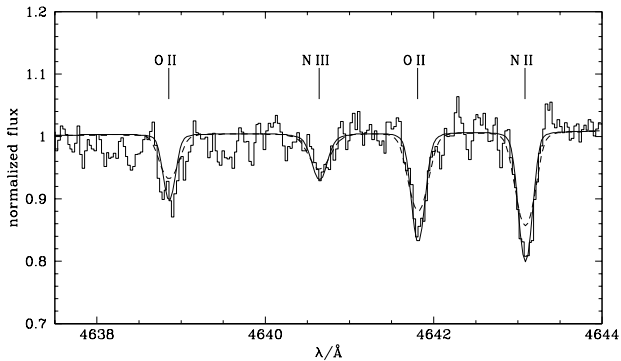}
\caption{Observed spectrum (coaddition of
the seven spectra obtained during best seeing) compared to synthetic spectra
with $v \sin i = 0$ (solid line) and $v \sin i = 10\,\mathrm{km\,s^{-1}}$
(dashed line) 
\label{f:vsini}}
\end{figure}

Following the procedure described in Heber et al.\ (\cite{HNR97}) we took
synthetic line profiles, convolved them with rotational profiles, and 
measured the fit quality (Fig.~\ref{f:vsini}). 
We limited this to lines of the ions
N\,II and O\,II because for these 
the quality of the atomic data is highest and these
lines are best reproduced by the model spectrum calculated with the 
abundances of Table~\ref{abu}.
A $3\sigma$ upper limit on $v \sin i$ 
of 4.7\,km\,s$^{-1}$ results. 

Since the synchronization time between 
orbit and rotation of the sdB is less than 700,000\,years
(Tassoul \& Tassoul \cite{TT92}), it is very likely
that orbital and rotational period are equal. This corresponds to a
rotational velocity of the sdB of 
$7.2\pm 0.8$\,km\,s$^{-1}$. Our upper limit on
$v \sin i$ thus transforms to an upper limit on the inclination angle of
this system of $48^\circ$. From the mass function we then derive a lower
limit on the mass of the invisible companion of $0.71\,M_\odot$. Thus the total
mass of the binary is $1.2\,M_\odot$ or even larger, close to the 
Chandrasekhar limit. A more definite determination would need dedicated 
observations with a narrow slit and a well defined spectral resolution.

\section{Discussion and Conclusions}

We measured accurate RVs for the sdB star He\,1047$-$0436
from high resolution spectra and derived an orbital period of 1.213253\,d 
and a semi-amplitude of 94\,km\,s$^{-1}$. The mass of the invisible 
companion 
is shown to be at least $0.44\,M_\odot$ indicating that the companion 
is likely a C/O core WD. If we make the reasonable assumption, that 
the sdB rotation is synchronized with the orbital period, we can use
an upper limit on $v \sin i$ to compute an upper limit on the inclination
angle and a lower limit on the WD mass of $0.71\,M_\odot$. 
Six of the seven known sdB+WD binaries (Maxted et al.\ \cite{max01}) 
have considerably shorter periods (0.09\,d to 0.57\,d) than HE\,1047$-$0436.
Only PG\,1538+269 has a longer orbital period (2.501\,d, Saffer et al.\ 
\cite{saf98}). Even the minimum mass we derive for the WD companion 
ranks third among the known systems.  

From a quantitative spectral analysis precise atmospheric parameters and 
metal abundances were derived. Intercomparison with an analysis of long 
slit spectra demonstrated that the atmospheric parameters derived 
from the Echelle spectra are highly reliable and systematic errors are 
small. The resulting atmospheric parameters and metal abundances are 
typical for sdB stars.

The effective temperature of HE\,1047$-$0436 lies within the range where short 
period (2--10\,min.) non-ra\-dial pul\-sa\-ting sdB stars were discovered 
(O'Do\-no\-ghue et al.\ \cite{odon99}). 
Pho\-to\-me\-tric monitoring of HE\,1047$-$0436 for about 
20\,min did not reveal any variations above a level of 2mmag
({\O}s\-ten\-sen, priv.\ com.)
\acknowledgement 
{We express our gratitude to the ESO staff, for providing invaluable help 
and conducting the service observations and pipeline reductions, which have 
made this work possible. C.~K., H.~E., and E.-M.~P.\ 
gratefully acknowledge financial support by the DFG (grants Na\,365/2-1
and He\,1354/30-1). 
This project was supported by DFG travel grant Na\,365/3-1.}

\end{document}